\documentclass[axioms,article,submit,moreauthors,pdflatex,10pt,a4paper]{mdpi-rev}

\firstpage{1}
\articlenumber{x}
\doinum{10.3390/------}
\pubvolume{xx}
\pubyear{2017}
\history{Received: date; Accepted: date; Published: date}

\usepackage{color}
\usepackage{amssymb,amsmath}
\usepackage{amssymb,pxfonts,txfonts}
\usepackage{bm}% bold math

\Title{Orientation Asymmetric Surface Model for Membranes : Finsler Geometry Modeling 
}

\Author{Evgenii Proutorov $^{1}$ and Hiroshi Koibuchi $^{2}$}
\AuthorNames{Evgenii Proutorov and Hiroshi Koibuchi}

\address{%
$^{1}$ \quad Cherepovets State University, Pr. Lunacharskii 5, 162600, Cherepovets, Russian Federation; proutorov@gmail.com\\
$^{2}$ \quad National Institute of Technology, Ibaraki College, Nakane 866, Hitachinaka, Ibaraki 312-8508, Japan}

\corres{Correspondence: koibuchi@mech.ibaraki-ct.ac.jp; koibuchih@gmail.com}

\simplesumm{We study triangulated surface models with nontrivial (=non-Euclidean) surface metrices for membranes. Our result shows that a nontrivial surface model is well-defined in the context of Finsler geometry modeling and the model is asymmetric under the surface inversion. }

\abstract{We study triangulated surface models with nontrivial surface metrices for membranes. The surface model is defined by a mapping  ${\bf r}$ from a two dimensional parameter space $M$  to the three dimensional Euclidean space ${\bf R}^3$. The metric variable $g_{ab}$, which is always fixed to the Euclidean metric $\delta_{ab}$, can be extended to a  more general  non-Euclidean metric on $M$ in the continuous model. The problem we focus on in this paper is whether such an extension is well-defined or not in the discrete model. We find that a discrete surface model with nontrivial metric becomes well-defined  if it is treated  in the context of Finsler geometry (FG) modeling, where triangle edge length in $M$ depends on the direction.  It is also shown that the discrete FG model is orientation assymetric on invertible surfaces in general, and for this reason, the FG model has a potential advantage for describing real physical membranes, which are expected to have some assymetries for orientation changing transformations.  
}


\begin{document}
%%%%%%%%%%%%%%%%%%%%%%%%%%%%%%%%%%%%%%%%%%
%% Only for the journal Gels: Please place the Experimental Section after the Conclusions

%%%%%%%%%%%%%%%%%%%%%%%%%%%%%%%%%%%%%%%%%%
%\setcounter{section}{-1} %% Remove this when starting to work on the template.
%\section{Introduction}
%The template details the sections that can be used in a manuscript. Note that the order of article sections may differ from the requirements of the journal (e.g. for the positioning of the Materials and Methods section). Please check the instructions for authors page of the journal to verify the correct order. For any questions, please contact the editorial office of the journal or support@mdpi.com. For LaTeX related questions please contact Janine Daum at latex-support@mdpi.com.

%%%%%%%%%%%%%%%%%%%%%%%%%%%%%%%%%%%%%%%%%%
%%%%%%%%%%%%%%%%%%%%%%%%%%%%%%%%%%%%%%%%%%
\section{Introduction}
%%%%%%%%%%%%%%%%%%%%%%%%%%%%%%%%%%%%%%%%%%
%%%%%%%%%%%%%%%%%%%%%%%%%%%%%%%%%%%%%%%%%%
%The introduction should briefly place the study in a broad context and highlight why it is important. It should define the purpose of the work and its significance. The current state of the research field should be reviewed carefully and key publications cited. Please highlight controversial and diverging hypotheses when necessary. Finally, briefly mention the main aim of the work and highlight the principal conclusions. As far as possible, please keep the introduction comprehensible to scientists outside your particular field of research. Citing a journal paper \cite{ref-journal}. And now citing a book reference \cite{ref-book}.
Biological membranes including artificial ones such as giant vesicles are simply understood as two-dimensional surfaces \cite{NELSON-SMMS2004}. The well-known surface model for membranes is statistical mechanically defined by using a mapping ${\bf  r}$ from a two-dimensional parameter space $M$ to ${\bf R}^3$ \cite{FDAVID-SMMS2004}. This mapping  ${\bf  r}$ and the metric $g_{ab} (a,b=1,2)$,  a set of functions on $M$, are the dynamical variables of the model.  To discretize these dynamical variables, we use triangulated surfaces in both $M$ and ${\bf R}^3$. On the discrete surfaces,  the metric $g_{ab}$ is always fixed to the Euclidean metric $\delta_{ab}$ \cite{KANTOR-NELSON-PRA1987,Bowick-PREP2001,GOMPPER-KROLL-SMMS2004}, while the induced metric $\partial_a {\bf r}\cdot\partial_b {\bf r} $  is also used in theoretical studies  on continuous surfaces \cite{FDAVID-SMMS2004}.  These  two-dimensional surface models are considered as a natural extension of one-dimensional polymer model \cite{Doi-Edwards-1986}, and a lot of studies for membranes have been conducted \cite{Lubensky-PRE-2003,HJHeermann-PRL2010,Lipowsky-SM2009,Noguchi-JPS2009,WIESE-PTCP19-2000}.  Landau-Ginzburg theory for membranes has also been developed \cite{PKN-PRL1988}. In Ref.  \cite{Koibuchi-Pol2016}, anisotropic morphologies of membranes  are studied, and the notion of multi-component is found to be essential also for the metric function \cite{GJug-PM2004}.  

However, it is still unclear whether non-Euclidean metric can be assumed or not for discrete models. In this  paper, we study the metric $g_{ab}$ in Ref.  \cite{Koibuchi-Pol2016} in more detail.  We will show that models with the metric in Ref.  \cite{Koibuchi-Pol2016} and their extension to a more general one are ill-defined in the ordinary surface modeling prescription, however, these ill-defined models turn to be well-defined  in the context of Finsler geometry (FG) modeling \cite{Koibuchi-Sekino-2014PhysicaA,George-Bogoslovsky-JGM2012,George-Bogoslovsky-PLA1998,Ootsuka-Tanaka-PLA2010,Matsumoto-SKB1975,Bao-Chern-Shen-GTM200}. Moreover,  it is also shown that the FG model becomes orientation asymmetric, where ''orientation asymmetric'' means that Hamiltonian is not invariant under the surface inversion \cite{Koibuchi-Pol2016}. In real physical membranes, the orientation asymmetry is observed because of their bilayer structure \cite{Seifert-PRE1994}. Indeed, asymmetry such as area difference between the outer and inner layers is expected to play an important role for anisotropic shape of membranes. Therefore, it is worth while to study the discrete surface model with non-trivial metric $g_{ab}$ more extensively. 

We should note that there are  two types of discrete surface models; the first is fixed connectivity (FC)  model and the second is dynamically triangulated (DT) surface model. The FC surface model corresponds to polymerized membranes, while the DT surface model corresponds to fluid membranes such as bilayer vesicles. The polymerized and fluid membranes are characterized by nonzero and zero shear moduli, respectively. Numerically, the dynamical triangulation for the DT models is simulated by bond-flip technique as one of the Monte Carlo processes  on triangulated lattices \cite{Ho-Baum-EPL1990,CATTERALL-PLB1989,Ambjorn-NPB1993},  while the FC surface models are defined on triangulated lattices without the bond flips. According to this classification, the discrete models in this paper belong to the DT surface models and correspond to fluid membranes, because the dynamical triangulation is assumed in the partition function, which will be defined in Section \ref{discrete-model},  just like in the model of \cite{Koibuchi-Pol2016}.

In Section \ref{continuous-model}, a continuous surface model and its basic properties are reviewed, and a non-Euclidean metric, which we study in this paper, is introduced. In Section \ref{discrete-model}, we discuss why orientation asymmetry needs to be studied, and then we introduce a discrete model on a triangulated spherical lattice and show that this discrete model is ill-defined in the ordinary context of surface modeling. In Section \ref{FG-model},  we show that this ill-defined model can be understood as a well-defined FG model in a modeling which is slightly extended from the one  in Ref. \cite{Koibuchi-Sekino-2014PhysicaA}. In Section \ref{discussion}, we summarize the results.

%%%%%%%%%%%%%%%%%%%%%%%%%%%%%%%%%%%%%%%%%%
%%%%%%%%%%%%%%%%%%%%%%%%%%%%%%%%%%%%%%%%%%
%\section{Results}
\section{Continuous surface model}\label{continuous-model}
%%%%%%%%%%%%%%%%%%%%%%%%%%%%%%%%%%%%%%%%%%
%%%%%%%%%%%%%%%%%%%%%%%%%%%%%%%%%%%%%%%%%%
%This section may be divided by subheadings. It should provide a concise and precise description of the experimental results, their interpretation as well as the experimental conclusions that can be drawn.

In this paper, we study a surface model which is an extension of the Helfrich and Polyakov (HP) model  \cite{HELFRICH-1973,POLYAKOV-NPB1986}. The HP model is physically defined by Hamiltonian $S$ which is a linear combination of the Gaussian bond potential $S_1$ and the bending energy $S_2$ such that
\begin{eqnarray} 
\label{cont_S1S2}
\begin{split}
&S=S_1+\kappa S_2,  \\
&S_1=\int \sqrt{g}d^2x g^{ab} \frac{\partial {\bf r}}{\partial x_a}\cdot \frac{\partial {\bf r}}{\partial x_b},  \\ 
&S_2=\frac{1}{2}\int \sqrt{g}d^2x  g^{ab} \frac{\partial {\bf n}}{\partial x_a} \cdot\frac{\partial {\bf n}}{\partial x_b}, 
\end{split}
\end{eqnarray} 
where $\kappa[k_BT]$ is the bending rigidity ($k_B$ and $T$ are the Boltzmann constant and the temperature, respectively). The surface position is described by ${\bf r}(\in {\bf R}^3)$, and $g_{ab}$ is a Riemannian metric on the two-dimensional surface $M$,  $g^{ab}\left(=\!(g_{ab})^{-1}\right)$ is its inverse, and $g\!=\!\det  g_{ab}$. Note that the surface position ${\bf r}$ is understood as a mapping ${\bf r}:M\ni x\!=\!(x_1,x_2)\mapsto \left(X(x),Y(x),Z(x)\right)\in {\bf R}^3$, where  the surface orientation is assumed to be preserved. The symbol ${\bf n}$ in $S_2$ denotes a unit normal vector of the image surface, where one of two orientations is used to define  ${\bf n}$.

It is well known that the Hamiltonian is invariant under (i) general coordinate transformation $x\!\to\! x^\prime$ in $M$ and (ii) conformal transformation for $g_{ab}$ such that $g_{ab}\!\to\! g_{ab}^\prime \!=\!f(x)g_{ab}$ with a positive function $f$ on $M$ \cite{FDAVID-SMMS2004}.  The first property under the transformation (i), called re-parametrization invariance, is expressed by $S({\bf r}(x), g_{ab}(x))\!=\!S({\bf r}(x^\prime), g_{ab}(x^\prime))$, where ${\bf r}(x^\prime)$ and $g_{ab}(x^\prime)$ are composite functions. The second property under (ii) is expressed by  $S({\bf r}(x), g_{ab}(x))\!=\!S({\bf r}(x), g_{ab}^\prime(x))$. The metrices $g_{ab}$ and $g^\prime_{ab}$ are called {\it conformally equivalent}, which is written as $g_{ab}\!\simeq\! g^\prime_{ab}$,  if there exists a positive function $f$ such that $g^\prime_{ab}\!=\!fg_{ab}$. Therefore, the second property with respect to the transformation (ii) implies that $S$ depends only on conformally non-equivalent metrices.

The metric $g_{ab}$ of the surface $M$ is generally given by $g_{ab}\!=\!\left(  
       \begin{array}{@{\,}cc}
        E &  F \\
        F &  G 
       \end{array} 
       \\ 
 \right)$ with the functions of $E\!>\!0,\; G\!>\!0, EG\!-\!F^2\!>\!0$. By letting $F\!=\!0$, we have $g_{ab}\!=\!\left(  
       \begin{array}{@{\,}cc}
        E &  0 \\
        0 &  G 
       \end{array} 
       \\  \right) \!=\! E\left(  
       \begin{array}{@{\,}cc}
        1 &  0 \\
        0 &  G/E 
       \end{array} 
       \\  \right)\!\simeq\! \left(  
       \begin{array}{@{\,}cc}
        1 &  0 \\
        0 &  \rho^2 
       \end{array} 
       \\  \right)\!\simeq\! \left(  
       \begin{array}{@{\,}cc}
        1/\rho &  0 \\
        0 &  \rho 
       \end{array} 
       \\  \right)$, where  $\rho^2\!=\!G/E$ \cite{Koibuchi-Pol2016}.  
This metric is in general not conformally equivalent to the  Euclidean metric $\delta_{ab}$. We call a metric $g_{ab}$ {\it trivial} ({\it non-trivial}) if $g_{ab}$ is conformally equivalent (inequivalent) to  $\delta_{ab}$, although  surface models with $g_{ab}\!=\!\delta_{ab}$ and $g_{ab}\!=\!\partial_a {\bf r}\cdot\partial_b {\bf r} $ are physically non-trivial \cite{Ho-Baum-EPL1990,CATTERALL-PLB1989,Ambjorn-NPB1993,DavidGuitter-1988EPL,NISHIYAMA-PRE-2004,KD-PRE2002,Kownacki-Mouhanna-2009PRE,Essa-Kow-Mouh-PRE2014,Cuerno-etal-2016}. 

%%%%%%%%%%%%%%%%%%%%%%%%%%%%%%%%%%%%%%%%%%
%%%%%%%%%%%%%%%%%%%%%%%%%%%%%%%%%%%%%%%%%%
\section{Discrete surface model}\label{discrete-model}
%%%%%%%%%%%%%%%%%%%%%%%%%%%%%%%%%%%%%%%%%%
%%%%%%%%%%%%%%%%%%%%%%%%%%%%%%%%%%%%%%%%%%
\subsection{Membrane orientation}\label{surface-orient}
%%%%%%%%%%%%%%%%%%%%%%%%%%%%%%%%%%%%%%%%%%
%=====================================
\begin{figure}[H]
\centering
\includegraphics[width=11.5cm]{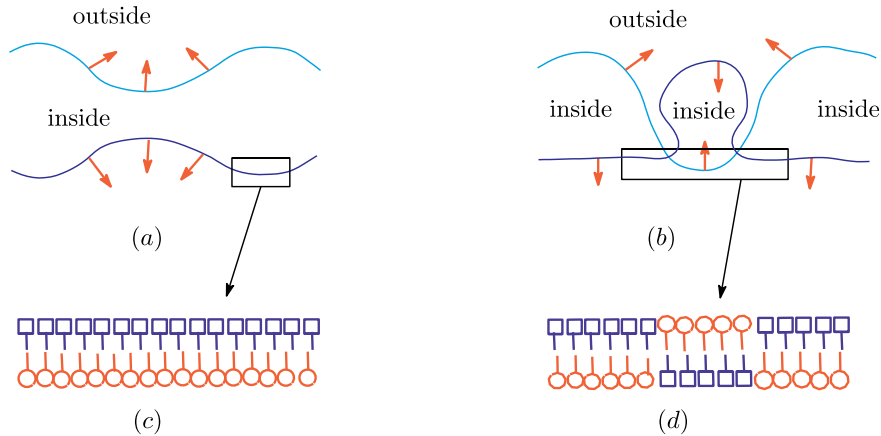}
\caption{A membrane in aqueous solution separates the solution into two regions; inside and outside.  (a) Self-avoiding surface with unit normal vectors ${\bf n}$, (b) self-intersecting surface with ${\bf n}$,  (c) lipid bilayer structure of membranes, where the symbols of lipids for inner and outer layers are drawn differently, and (d) a partly inverted bilayer.  }
\label{fig-1}
\end{figure}
%=====================================
First, we should comment on the surface orientation. The unit normal vector ${\bf n}$ is directed from inside to outside of material separated from bulk material by membrane (see Fig. \ref{fig-1}(a)). 
However, if the membrane self-intersects, then the direction of ${\bf n}$ changes from outside to inside (Fig. \ref{fig-1}(b)). Otherwise ($\Leftrightarrow$ ${\bf n}$ is directed from inside to outside), ${\bf n}$ discontinuously changes at the intersection point. For this reason, we change the surface orientation by changing the local coordinate system from left-handed to right-handed while ${\bf n}$ remains unchanged (Fig. \ref{fig-1}(b)).  We should emphasize that our basic assumption is that the surface orientation is locally changeable. This means that the surface in ${\bf R}^3$ is self-intersecting, or in other words the surface is not self-avoiding.

However, such intersection process is not so easy to implement in the numerical simulations (no numerical simulation is performed in this paper). Apart from this, it is unclear whether  or not the implementation of such intersection process is effective for simulating the membrane inversion. 
Therefore, we assume that the surface is locally invertible without intersections;  an inversion is expected to occur independent of whether the surface is self-intersecting or not. Indeed, real physical membranes  are composed of lipid molecules, which have hydrophobic and hydrophilic parts. These lipids form a bilayer structure (Fig. \ref{fig-1}(c)). 
In those real membranes, the bilayer structure is partly inverted just as in Fig. \ref{fig-1}(d) via the so-called flip-flop process. Such inversion process without intersection is not always unphysical because it can be seen in the process of pore formation. The pore formation process is reversible and forms cup-like membranes, where the membranes are not always self-intersecting \cite{Hotani-PNAS1998}.  The cup-like membranes are stable \cite{Suezaki-JCPB2002} and expected to play an important role as an intermediate configuration for cell inversion. It should be remarked that the surface orientation is also changeable in the process of cell fission and fusion, where the surface self-intersects, in real physical membranes. 
 
\begin{figure}[H]
\centering
\includegraphics[width=12.5cm]{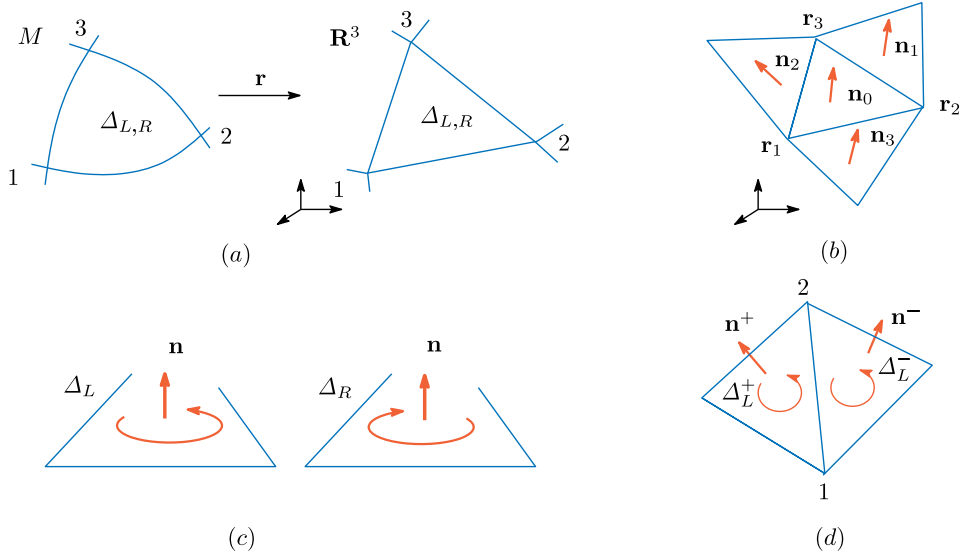}
\caption{(a) A mapping ${\bf r}$ from a smooth triangle in $M$ to a piecewise linear triangle in ${\bf R}^3$, (b) a triangle and the three neighboring triangles in ${\bf R}^3$, (c) the definition of ${\bf n}$ on the triangles ${\it \Delta}_{L,R}$,  and (d) two neighboring triangles ${\it \Delta}^\pm_L$ with unit normal vectors ${\bf n}^\pm$ and their common bond $12$. The suffices of ${\it \Delta}_{L,R}$  denote the orientation of the triangle. The open circles with an arrow at one terminal point indicate the surface orientation. }
\label{fig-2}
\end{figure}

 To define a discrete model, we use a piecewise-linearly triangulated surface in ${\bf R}^3$ \cite{KANTOR-NELSON-PRA1987,Bowick-PREP2001,GOMPPER-KROLL-SMMS2004}. In this paper, a spherical surface is assumed. Therefore, it is natural to assume that $M$ is also triangulated and of sphere topology. Triangles in $M$ can be smooth in general, and these smooth triangles are mapped to piecewise-linear triangles in ${\bf R}^3$ by ${\bf r}$ (see Figs. \ref{fig-2}(a) and \ref{fig-2}(b))). We should note that triangle ${\it \Delta}$ in $M$ has two different orientations. Let ${\it \Delta}_{L,R}$ denote the triangle that has the left-hand (right-hand) orientation,  where $L(R)$ corresponds to the left-handed (right-handed) local coordinate system. The symbol ${\it \Delta}_{L}$ is used for non-inverted parts of the surface, while  ${\it \Delta}_{R}$  is used for inverted parts shown in Fig. \ref{fig-1}(d).  The direction of ${\bf n}$ is defined to be dependent on the orientation of ${\it \Delta}_{L,R}$ as mentioned in the previous subsection (see Fig. \ref{fig-2}(c)).

%=====================================
%=====================================
The surface inversion is given by 
\begin{equation}
\label{inversion}
{\bf r}_i \to -{\bf r}_i \quad ({\rm for \; all}\; i),
\end{equation} 
for example. The problem is whether the inverted surface is stable or not. As we will see below, the  energy of the inverted surface is different from that of the original surface in a non-Euclidean metric model. This non-Euclidean metric model becomes well-defined if it is treated as an FG model.  In the FG modeling (not in the standard HP modeling), we assume that the surface is locally invertible as in Fig. \ref{fig-1}(d), which can be defined by  the change of local coordinate orientation. Thus, studies on the stability  of inverted surfaces become feasible within the scope of FG modeling, although the transformation of variables ${\bf r}_i$ for this local inversion  is not always given by Eq. (\ref{inversion}); the vertex position remains unchanged under the change of triangle orientation.   

%%%%%%%%%%%%%%%%%%%%%%%%%%%%%%%%%%%%%%%%%%
\subsection{Discretization of the model}\label{discretization}
%%%%%%%%%%%%%%%%%%%%%%%%%%%%%%%%%%%%%%%%%%
In this subsection, the discretization of Hamiltonian in Eq. (\ref{cont_S1S2}) is performed on the triangles  ${\it \Delta}_L$ and their image triangles ${\bf r}\left({\it \Delta}_L\right)$. 
The function $\rho$ in $g_{ab}$ is defined on each triangle ${\it \Delta}$ in $M$ in the discrete model, and we denote the function $\rho$ on ${\it \Delta}$ by $\rho_{\it \Delta}$. 
Thus, the discrete metric defined on triangle ${\it \Delta}$ is given by 
\begin{equation}
\label{org_cont_metric}
g_{ab}=\left(  
       \begin{array}{@{\,}cc}
        1/\rho_{\it \Delta} &  0 \\
        0 &  \rho_{\it \Delta} 
       \end{array} 
       \\ 
 \right), \quad \rho_{\it \Delta}>0, \quad ({\rm on}\; {\it \Delta}_L).
\end{equation} 
By replacing the integral and partial derivatives in $S_1$ and $S_2$ with  the sum over triangles ${\it \Delta}$ and differences, respectively, such that
\begin{eqnarray} 
\label{discretization}
\begin{split}
&\int \sqrt{g}d^2x \to \sum_{\it \Delta}, \\
&\frac{\partial {\bf r}}{\partial x_1}\to{\bf r}_2-{\bf r}_1,\quad \frac{\partial {\bf r}}{\partial x_2}\to{\bf r}_3-{\bf r}_1,\\
&\frac{\partial {\bf n}}{\partial x_1}\to{\bf n}_0-{\bf n}_2,\quad \frac{\partial {\bf n}}{\partial x_2}\to{\bf n}_0-{\bf n}_3,
\end{split}
\end{eqnarray} 
we have the discrete expressions  $g^{11}({\bf r}_2\!-\!{\bf r}_1)^2\!+\! g^{22}({\bf r}_3\!-\!{\bf r}_1)^2$ and  $g^{11}({\bf n}_0\!-\!{\bf n}_2)^2\!+\! g^{22}({\bf n}_0\!-\!{\bf n}_3)^2$ corresponding to the discrete energies $g^{ij}({\partial {\bf r}}/{\partial x_i})\cdot({\partial {\bf r}}/{\partial x_j})$ and $g^{ij}({\partial {\bf n}}/{\partial x_i})\cdot({\partial {\bf n}}/{\partial x_j})$ of $S_1$ and $S_2$ on triangle ${\it \Delta}$, where the local coordinate origin is assumed at vertex $1$ (see Fig. \ref{fig-2}(b)).  Thus, the corresponding discrete expressions of $S_1$ and $S_2$ are given by 
\begin{eqnarray}
\label{Disc-Eneg-model-1}
\begin{split}
&S_1=\sum_{\it \Delta}S_1\left({\it \Delta}\right)=\sum_{\it \Delta}\left(\rho \ell_{12}^2+\frac{1}{\rho}\ell_{13}^2\right),\\ 
&S_2=\sum_{\it \Delta}S_2\left({\it \Delta}\right)=\sum_{\it \Delta} \left[\rho\left(1-{\bf n}_0\cdot{\bf n}_2\right)+\frac{1}{\rho} \left(1-{\bf n}_0\cdot{\bf n}_3\right)\right],
\end{split}
\end{eqnarray}
where $\ell_{ij}=|\vec \ell_{ij}|=|{\bf r}_j-{\bf r}_i|$.  
The index $i$ of ${\bf n}_i$ in this $S_2$ represents a triangle (see Fig. \ref{fig-2}(b)). Since the coordinate origin can also be assumed at vertices $2$ and $3$ on triangle ${\it \Delta}$, we have three possible discrete expressions including those in Eq. (\ref{Disc-Eneg-model-1}) for $g^{ij}({\partial {\bf r}}/{\partial x_i})\cdot({\partial {\bf r}}/{\partial x_j})$ and $g^{ij}({\partial {\bf n}}/{\partial x_i})\cdot({\partial {\bf n}}/{\partial x_j})$. Thus, we have:
\begin{eqnarray}
\label{Disc-Eneg-model-2}
\begin{split}
&S_1=\frac{1}{3}\sum_{\it \Delta} \left[\left(\rho_{1}+\frac{1}{\rho_{2}}\right) \ell_{12}^2+\left(\rho_{2}+\frac{1}{\rho_{3}}\right)\ell_{23}^2+\left(\rho_{3}+\frac{1}{\rho_{1}}\right)\ell_{31}^2\right],\\ 
&S_2=\frac{1}{3}\sum_{\it \Delta} \left[\left(\rho_{2}+\frac{1}{\rho_{1}}\right) \left(1-{\bf n}_0\cdot{\bf n}_3\right)+\left(\rho_{3}+\frac{1}{\rho_{2}}\right)\left(1-{\bf n}_0\cdot{\bf n}_1\right) \right.\\
&\qquad +\left.\left(\rho_{1}
+\frac{1}{\rho_{3}}\right)\left(1-{\bf n}_0\cdot{\bf n}_2\right)\right],
\end{split}
\end{eqnarray}
where the factor $1/3$ is assumed. In the expressions, the suffix $i$ of $\rho_i$ denotes the coordinate origin. The reason why the function $\rho$ depends on the coordinate origin is that $\rho$ is an element of $2\times 2$ matrix $g_{ab}$, which depends on local coordinates in general. 

The expressions for $S_1$ and $S_2$ in Eqs. (\ref{Disc-Eneg-model-1}) and (\ref{Disc-Eneg-model-2}) correspond to those for ${\it \Delta}_L$. In Eq. (\ref{Disc-Eneg-model-2}), the sum over triangles $\sum_{\it \Delta}$ in $S_1$ and $S_2$ can be replaced by sum over bonds $\sum_{ij}$. In this replacement, we should remind ourselves of the fact that the first terms of $S_1$ and $S_2$ in  (\ref{Disc-Eneg-model-2}) are respectively replaced by $\left(\rho^+_{1}\!+\!{1}/{\rho^+_{2}}\!+\!\rho^-_{2}\!+\!{1}/{\rho^-_{1}}\right) \ell_{12}^2$ and $\left(\rho^+_{2}\!+\!{1}/{\rho^+_{1}}\!+\!\rho^-_{1}\!+\!{1}/{\rho^-_{2}}\right) \left(1\!-\!{\bf n}^+\cdot{\bf n}^-\right)$. In these expressions, $\rho^\pm_i$ denotes the function $\rho$ on the triangles ${\it \Delta}^\pm_L$,  where the coordinate origin is at vertex $i$ (see Fig. \ref{fig-2}(d)), and ${\bf n}^\pm$ denote  ${\bf n}$ for triangles ${\it \Delta}^\pm_L$. The coefficient of $\ell_{12}^2$ is different from that of $\left(1\!-\!{\bf n}^+\cdot{\bf n}^-\right)$, and these coefficients come from the following expressions:
\begin{eqnarray}
\label{Disc-Enegs} 
\begin{split}
&S_1({\it \Delta}^+_L) =\left(\rho^+_{1}+{1}/{\rho^+_{2}}\right)\ell_{12}^2+\cdots,\\
&S_1({\it \Delta}^-_L)=\left(\rho^-_{2}+{1}/{\rho^-_{1}}\right)\ell_{12}^2+\cdots, \\ 
&S_2({\it \Delta}^+_L) =\left(\rho^+_{2}+{1}/{\rho^+_{1}}\right)\left(1-{\bf n}^+\cdot{\bf n}^-\right)+\cdots,\\
&S_2({\it \Delta}^-_L) =\left(\rho^-_{1}+{1}/{\rho^-_{2}}\right)\left(1-{\bf n}^+\cdot{\bf n}^-\right)+\cdots.  
\end{split}
\end{eqnarray}
Thus, we have
\begin{eqnarray}
\label{Disc-Eneg-model-3} 
\begin{split}
&S_1=\sum_{ij} \gamma_{ij}\ell_{ij}^2, \quad S_2=\sum_{ij} \kappa_{ij}\left(1-{\bf n}^+\cdot{\bf n}^-\right), \\
& \gamma_{ij} =\left(\gamma_{ij}^++\gamma_{ij}^-\right)/4, \quad \kappa_{ij}= \left(\kappa_{ij}^++\kappa_{ij}^-\right)/4,\\
&\gamma_{ij}^+= \rho^+_{i}+{1}/{\rho^+_{j}}, \quad \kappa_{ij}^+=\rho^+_{j}+{1}/{\rho^+_{i}},\quad ({\rm on}\;{\it \Delta}^+_L), \\
 &\gamma_{ij}^-= \rho^-_{j}+{1}/{\rho^-_{i}},  \quad \kappa_{ij}^-=\rho^-_{i}+{1}/{\rho^-_{j}},\quad  ({\rm on}\; {\it \Delta}^-_L), \\
 \end{split}
\end{eqnarray}
where the factor $1/3$ is replaced by $1/4$ in the final expressions of $S_1$ and $S_2$. The indices $ij$ of   $\gamma_{ij}$ and $\kappa_{ij}$  simply denote vertices $i$ and $j$. We should note that $\gamma_{ij} \!=\!\kappa_{ji}$ and $\gamma_{ij} \!\not=\!\kappa_{ij}$ in general in Eq. (\ref{Disc-Eneg-model-3}) as mentioned above. 

The partition function $Z$ and Hamiltonian $S$ of the model, we start with in this paper, are defined by 
\begin{eqnarray} 
\label{Part-Func-flu}
\begin{split}
&Z(\lambda,\kappa) =  \sum_\sigma \sum_{\mathcal T}\int^\prime \prod _{i=1}^{N} d {\bf r}_i \exp\left[-S({\bf r},\sigma)\right],\\
&S=\lambda S_0+S_1+\kappa S_2,\quad S_0=\sum_{\pm}\left(1-\sigma^+\cdot\sigma^-\right), \quad \left(\sigma^\pm\in \{1,-1\}\right),
\end{split}
\end{eqnarray} 
where Ising model energy $S_0$ with the coefficient $\lambda$ is included in $S$. This is a surface model for multi-component membranes \cite{Koibuchi-Pol2016}. The sum $\sum_{\pm}$ in $S_0$ denotes the sum over all nearest neighbor triangles $+$ and $-$, and $\sigma^\pm$ denotes that $\sigma$ is defined on the triangles ${\it \Delta}^\pm_L$.  The variable $\sigma$ is an element of ${\bf Z}_2\!=\!\{1,-1\}$, however, $S_0$ (and $\sigma$) is not always limited to Ising type Hamiltonian. The variable $\sigma^\pm$ is introduced to represent the components A and B such as liquid-ordered and liquid-disordered phases \cite{Koibuchi-Pol2016}.  If $\sigma^+\!=\!1 (-1)$ on triangle ${\it \Delta}^+$, this triangle ${\it \Delta}^+$ is understood such that it belongs to or is occupied by the component A (B) for example. The value of $\sigma$ on each triangle ${\it \Delta}$ remains unchanged, however, the energy $S_0$ does not remain constant because the combination of nearest neighbor pairs of triangles ${\it \Delta}^\pm$ changes due to the triangle diffusion, which is actually expected on dynamically triangulated surfaces \cite{Koibuchi-Pol2016}. In the model of Ref. \cite{Koibuchi-Pol2016}, the function $\rho_i^+$  is independent of vertex $i$ and depends only on  triangle ${\it \Delta}^+$,  and therefore the value of $\rho^+$ is uniquely determined only by $\sigma^+$ if the  dependence of  $\rho^+$ on $\sigma^+$ is fixed. As a consequence, the metric $g_{ab}$ is determined by the internal variable $\sigma$.  In the model of Eq. (\ref{Part-Func-flu}), the dependence of $\rho_i^+$ on $\sigma^+$ is not explicitly specified, because this dependence of $\rho$ on $\sigma$ is in general independent of the well definedness of discrete surface models with non-Euclidean metric, and this well definedness is the main target in this paper. 

In $Z$, $\sum_\sigma$ and $\sum_{\mathcal T}$ denote the sum over all possible configurations of $\sigma$ and triangulations ${\mathcal T}$, respectively. The sum over triangulation $\sum_{\mathcal T}$ can be simulated by the bond flips in MC simulations, and therefore the model is grouped into the fluid surface models as mentioned in the Introduction. The symbol ${\mathcal T}$ in  $\sum_{\mathcal T}$ denotes the triangulation, which is assumed as one of the dynamical variables of the discrete fluid model. This means that a variable ${\mathcal T}$ corresponds to a triangulated lattice configuration. Therefore, the lattice configurations in the parameter space $M$ are determined by  ${\mathcal T}$. On the other hand a lattice configuration corresponding to a given ${\mathcal T}$ is originally considered as an ingredient of a set of local coordinate systems; two different ${\mathcal T}$s correspond to two inequivalent coordinates which are not transformed to each other by any coordinate transformation. Recalling that the continuous Hamiltonian is invariant under general coordinated transformations, we can chose an arbitrary coordinate such as orthogonal coordinate for each triangle of a given  ${\mathcal T}$. However, from the Polyakov's string theoretical point of view, the partition function is defined by the sum over all possible metrices $\int {\mathcal D}g$ in addition to the sum over all possible mappings $\int {\mathcal D}{\bf r}$. Since the metric $g$ depends on coordinates,  $\int {\mathcal D}g$ is considered to be corresponding to the sum over local coordinates, which is simulated by $\sum_{\mathcal T}$ in the discrete models. Therefore, from these intuitive discussions, the Euclidean metric, for example, is forbidden in a fluid model on triangulated lattices without DT; this Euclidean metric model without DT is simply a FC model for polymerized membranes, where the surface inversion is not expected.

The symbol $\int^\prime \prod _{i=1}^{N} d {\bf r}_i $ denotes $3(N-1)$-dimensional integrations in ${\bf R}^3$ under the condition that the center of mass of the surface is fixed to the origin of ${\bf R}^3$. The Hamiltonian $S$ has the unit of energy $[k_BT]$. The coefficient $\kappa[k_BT]$ of $S_2$ is the bending rigidity.

Here, we comment on the property called {\it scale invariance} of the model \cite{WHEATER-JP1994}. This comes from the fact that the integration of ${\bf r}$ in $Z$ is independent of the scale transformation such that ${\bf r\!}\to\! \alpha {\bf r}$ for arbitrary positive $\alpha\!\in\! {\bf R}$. This property is expressed by  $Z(\{{\bf r}\})\!=\!Z(\{\alpha{\bf r}\})$, and therefore,  for Hamiltonian $S^\prime\!=\!\lambda S_0\!+\!cS_1\!+\!\kappa S_2$, we have 
\begin{eqnarray} 
\label{Part-Func-scale-inv}
\begin{split}
&\sum_\sigma \sum_{\mathcal T}\int^\prime \prod _{i=1}^{N} d {\bf r}_i \exp\left[-S^\prime({\bf r})\right]\\
=&\alpha^{3N-1}\sum_\sigma \sum_{\mathcal T}\int^\prime \prod _{i=1}^{N} d {\bf r}_i \exp\left[-\left(\lambda S_0+c\alpha^2 S_1+\kappa S_2\right)\right]\\
=&c^{-(3N-1)/2}\sum_\sigma \sum_{\mathcal T}\int^\prime \prod _{i=1}^{N} d {\bf r}_i \exp\left[-\left(\lambda S_0+S_1+\kappa S_2\right)\right]\\
=&c^{-(3N-1)/2}\sum_\sigma \sum_{\mathcal T}\int^\prime \prod _{i=1}^{N} d {\bf r}_i \exp\left[-S({\bf r})\right].
\end{split}
\end{eqnarray} 
In the second line of Eq. (\ref{Part-Func-scale-inv}), we assume $\alpha\!=\!1/\sqrt{c}$, and then in the third line  we have $S^\prime(\alpha {\bf r})\!=\!\lambda S_0\!+\!S_1\!+\!\kappa S_2$ because $S_0$ and $S_2$ are scale independent and $S_1(\alpha {\bf r})\!=\!\alpha^2S_1({\bf r})$.  
Thus, from the fact that the partition function is independent of multiplicative constant, we find that the model with  $S^\prime\!=\!\lambda S_0\!+\!cS_1\!+\!\kappa S_2$ is equivalent to the model with $S\!=\!\lambda S_0\!+\!S_1\!+\!\kappa S_2$.  "Equivalent" means that the shape of surface is independent of the value of $c(\!>\!0)$ although the surface size depends on $c$ in general. The dependence of surface size on $c$ is also understood from the scale invariant property of $Z$. Indeed,  it follows from $Z(\{{\bf r}\})\!=\!Z(\{\alpha{\bf r}\})$ that $\partial Z(\{\alpha{\bf r}\})/\partial \alpha |_{\alpha=1}\!=\!0$, and therefore we have \cite{WHEATER-JP1994}
\begin{eqnarray} 
\label{Part-Func-scale-dep}
\begin{split}
&\left. \frac{\partial \log Z\left[S^\prime(\alpha{\bf r})\right]}{\partial \alpha} \right|_{\alpha=1} \\
=&\frac{1}{Z}\left[(3N-1) \alpha^{3N-2} Z -2c \alpha^{3N}\sum_\sigma \sum_{\mathcal T}\int^\prime \prod _{i=1}^{N} d {\bf r}_i S_1\exp\left[-\left(\lambda S_0+c\alpha^2 S_1+\kappa S_2\right)\right]  \right]_{\alpha=1} \\
=&(3N-1) - 2c\langle S_1\rangle=0  \quad \Leftrightarrow \quad \langle S_1\rangle/N= {3}/({2c}).
\end{split}
\end{eqnarray} 
This final equation implies that the mean bond length squares $\langle \ell_{ij}^2\rangle$ depends on $c$, because $S_1$ is given by $S_1\!=\!\sum_{ij} \gamma_{ij} \ell_{ij}^2$ where $\gamma_{ij}$ is independent of $c$.  For a specialized case that $\gamma_{ij}$=constant, $\langle S_1\rangle$ becomes proportional to $\langle \ell_{ij}^2\rangle$. On the other hand, the mean bond length squares in general represent the surface size for smooth surfaces, which are expected for sufficiently large $\kappa$. 

We should note that the model studied in Ref. \cite{Koibuchi-Pol2016} for a two-component membrane is obtained from the model of Eqs. (\ref{Disc-Eneg-model-3}) and (\ref{Part-Func-flu}) by the assumption that $\rho^\pm_i$ is independent of the local coordinate origin $i$ and depends only on  triangles ${\it \Delta}^\pm$. In this case, the model is orientation symmetric, and therefore the lower suffices $L,R$ for the orientation of triangles ${\it \Delta}_{L,R}$ are not necessary. Then, we have $\gamma_{ij}\!=\!\kappa_{ij}\!=\!({1}/{4})\left(\rho^+\!+\!{1}/{\rho^+}\!+\!\rho^-\!+\!{1}/{\rho^-}\right)$, where $+$ and $-$ are the two neighboring triangles of  bond $ij$ which links vertices $i$ and $j$.  Thus,  $\gamma_{ij}$ (and $\kappa_{ij}$) defined on bond $ij$ depends only on $\rho^\pm$ of the two neighboring triangles in the model of Ref. \cite{Koibuchi-Pol2016}.  For this reason, the configuration (or distribution) of $\rho$ on the surface remains unchanged if the triangulation is fixed. However, the model is defined on dynamically triangulated lattices, which allow not only vertices but also triangles to diffuse freely over the surface \cite{Ho-Baum-EPL1990,CATTERALL-PLB1989,Ambjorn-NPB1993}. This free diffusion of triangles changes the distribution of $\rho$ and hence  $\gamma_{ij}$ and $\kappa_{ij}$.  Moreover, $\sigma (\in {\bf Z}_2)$ is assigned on triangles (not on vertices) such that the value of $\rho^\pm$ on each triangle is determined by $\sigma^\pm (\in {\bf Z}_2)$. As a consequence, the corresponding energy $S_0\!=\!\sum_{\pm}\left(1\!-\!\sigma^+\cdot\sigma^-\right)$ becomes dependent on the distribution of $\rho$, or in other words, the distribution of  $\gamma_{ij}$ and $\kappa_{ij}$ is determined by the energy $S_0$.    This is an outline of the model in Ref. \cite{Koibuchi-Pol2016}. 

In this paper,  $\rho^\pm_i$ depends on not only triangles ${\it \Delta}^\pm$ but also  the local coordinate origin $i$ in contrast to that of the model in Ref. \cite{Koibuchi-Pol2016}. We should note that the relation between $\rho^\pm$ and $\sigma^\pm$ is not explicitly specified. Although the model is not determined without the explicit relation,  the following discussions in this paper are independent of this relation.  

%%%%%%%%%%%%%%%%%%%%%%%%%%%%%%%%%%%%%%%%%%
\subsection{Well-defined model}
%%%%%%%%%%%%%%%%%%%%%%%%%%%%%%%%%%%%%%%%%%
We start with the definition of {\it trivial} ({\it non-trivial})  model for a discrete surface model.
\begin {Definition}
Let us assume that Hamiltonian $S$ of a discrete surface model is given by Eq. (\ref{Disc-Eneg-model-3}).  
Then, this discrete model is called trivial (non-trivial) if the following conditions are (not) satisfied:
\begin{eqnarray} 
\label{trivial-condition}
\gamma_{ij}={\rm constant},\quad \kappa_{ij}={\rm constant},
\end{eqnarray} 
where the constants are independent of bond $ij$, and these constants are not necessarily be the same.
\end {Definition}

We assume $\lambda\!=\!0$ in $S$ of Eq. (\ref{Disc-Eneg-model-3}) for simplicity.  We should note that a model with $S^\prime\!=\!c_1S_1\!+\!\kappa c_2S_2$, for arbitrary coefficients $c_1$ and $c_2$, is identical to the model defined by $S\!=\!S_1\!+\!\kappa^\prime S_2$ with $\kappa^\prime\!=\!\kappa c_2$. Indeed, because of the scale invariance of $Z$ discussed in the previous subsection using Eq. (\ref{Part-Func-scale-inv}), the coefficient $c_1$ of $S_1$ in  $S^\prime$ can be replaced by $1$. Thus, we have $S^\prime\!=\!S_1\!+\!\kappa^\prime S_2$.

If the metric is conformally equivalent to Euclidean metric, then the model is trivial.  In this sense, this definition for trivial (non-trivial) model is an extension of the definition by the terminology  {\it conformally equivalent} for $g_{ab}$ discussed in Section \ref{surface-orient}. However, there exists a metric, that is conformally non-equivalent  to $\delta_{ab}$ while it  makes the model trivial. An example of such metric is $g_{ab}\!=\!\left(  
       \begin{array}{@{\,}cc}
        2/(3\!+\!\sqrt{5}) & 0 \\
        0 &  (3\!+\!\sqrt{5})/2
       \end{array} 
       \\ 
        \right)$, and more detailed information will be given below (in Remark 2). 

%=====================================
\begin{figure}[H]
\centering
\includegraphics[width=12cm]{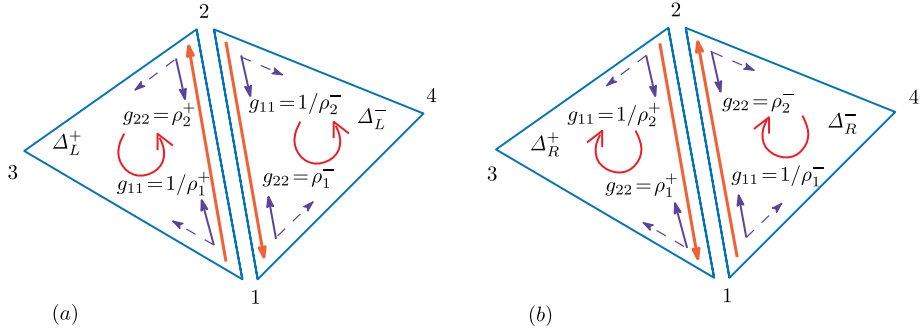}
\caption{(a) Two neighboring triangles ${\it \Delta}^\pm$  and elements of $g_{ab}$ for the direction dependent length of bond $12$ at the vertices $1$ and $2$, and (b) the inverted triangles  ${\it \Delta}^\pm_R$  (inside view) of ${\it \Delta}^\pm_L$ in (a). The direction dependent lengths of bond $12$ are indicated by long arrows in both ${\it \Delta}^\pm_L$ and ${\it \Delta}^\pm_R$. }
\label{fig-3}
\end{figure}
%=====================================
Next, we introduce the notion of {\it direction dependent length} $L_{ij}$ (and $L_{ji}$) of bond $ij$, which is shared by two triangles, in the discrete model. Let  ${\it \Delta}^\pm_L$ be  the two nearest neighbor triangles of bond $12$ on $M$ (Fig.\ref{fig-3}(a)). The length $L_{12}({\it \Delta}^+_L)$ of bond $12$  is defined by  $L_{12}({\it \Delta}^+_L)\!=\!\int dx_11/ \rho^+_1\!=\!1/\rho^+_1$, where $ 1/\rho^+_1$ is the element  $g_{11}$ of the metric $g_{ab}$ on ${\it \Delta}^+_L$ where the local coordinate origin is at vertex $1$; the symbol ${\it \Delta}^+_L$ in $L_{12}({\it \Delta}^+_L)$ denotes that $L_{12}$ is defined by $g_{ab}$ on triangle ${\it \Delta}^+_L$. It is also possible to define $L_{12}({\it \Delta}^+_L)$ by  $L_{12}({\it \Delta}^+_L)\!=\!\int dx_2(\rho^+_2)\!=\!\rho^+_2$, where $ \rho^+_2$ is the element  $g_{22}$ on ${\it \Delta}^+_L$ where the local coordinate origin is at vertex $2$. Thus,  $L_{12}({\it \Delta}^+_L)$ is defined by the mean value of these two lengths, and the length $L_{21}({\it \Delta}^-_L)$ of bond $12$ is also defined with exactly same manner. Then, we have 
\begin{eqnarray} 
\label{bond-length}
L_{12}({\it \Delta}^+_L)=(1/2)\left(1/\rho^+_1+\rho^+_2\right),\quad L_{21}({\it \Delta}^-_L)=(1/2)\left(1/\rho^-_2+\rho^-_1\right).
\end{eqnarray} 
These two lengths are different from each other in their expressions, and therefore it appears that the bond length is dependent on its direction. 
For the inverted surface (shown in  Fig.\ref{fig-3}(b)),  we also have the two different lengths 
\begin{eqnarray} 
\label{bond-length-invert}
\bar L_{12}({\it \Delta}^-_R)=(1/2)\left(1/\rho^-_1+\rho^-_2\right), \quad\bar L_{21}({\it \Delta}^+_R)=(1/2)\left(1/\rho^+_2+\rho^+_1\right). 
\end{eqnarray} 

It is also possible to define the lengths of bond $12$ as follows:
\begin{eqnarray} 
\label{bond-length-prime}
\begin{split}
&L_{12}^\prime({\it \Delta}_L)=(1/2)\left(1/\rho^+_1+\rho^-_1\right),\quad L_{21}^\prime({\it \Delta}_L)=(1/2)\left(\rho^+_2+1/\rho^-_2\right),\\
&\bar L_{12}^\prime({\it \Delta}_R)=(1/2)\left(\rho^+_1+1/\rho^-_1\right), \quad\bar L_{21}^\prime({\it \Delta}_R)=(1/2)\left(1/\rho^+_2+\rho^-_2\right), 
\end{split}
\end{eqnarray}
where $L_{12}^\prime$ and $L_{21}^\prime$ ($\bar L_{12}^\prime$ and $\bar L_{21}^\prime$) correspond to those in Eq. (\ref{bond-length}) (Eq. (\ref{bond-length-invert})). The following discussions remain unchanged if $L_{12}^\prime$, $L_{21}^\prime$ and $\bar L_{12}^\prime$, $\bar L_{21}^\prime$ are assumed as the definition of bond lengths. For this reason, we use only the expressions in Eq. (\ref{bond-length}) and Eq. (\ref{bond-length-invert}) for bond lengths in the discussions below.

Now, let us introduce the notion of  {\it well-defined model}. 
\begin {Definition}
A discrete surface model is called well-defined if the following conditions are satisfied:
\begin{enumerate}[leftmargin=1cm,labelsep=3mm]
\item[{\rm (A1)}]	Any bond length is independent of its direction
\item[{\rm (A2)}]	Any bond length is independent of surface orientation
\item[{\rm (A3)}]	Any triangle area is independent of surface orientation
\end{enumerate}
\end {Definition}
We should note that these constraints (A1)--(A3) are not imposed on Finsler geometry models, which will be introduced in the following section. 
Using Eqs. (\ref{bond-length}) and (\ref{bond-length-invert}), we rewrite the first and second conditions (A1) and (A2) such that 
\begin{eqnarray} 
\label{expression-definition-1-1}
&& 1/\rho^+_1+\rho^+_2=1/\rho^-_2+\rho^-_1, \quad 
%1/\rho^-_1+\rho^-_2=1/\rho^+_2+\rho^+_1 
\quad (\Leftrightarrow {\rm (A1)}), \\
\label{expression-definition-1-2}
&& 1/\rho^-_2+\rho^-_1=1/\rho^-_1+\rho^-_2, \quad 
1/\rho^+_1+\rho^+_2=1/\rho^+_2+\rho^+_1 \quad (\Leftrightarrow {\rm (A2)}). 
\end{eqnarray} 
 The condition (A3) is always satisfied because of the fact that $\det g_{ab}\!=\!1$ for the metric function in Eq. (\ref{org_cont_metric}).  Note that the constraint (A1) is imposed only on triangles $({\it \Delta}_L$, and the equation corresponding to (A1) on triangles $({\it \Delta}_L$ is not independent of the three equations in Eqs. (\ref{expression-definition-1-1}) and  (\ref{expression-definition-1-2}).

If we use the following definition for the bond length consistency for every vertex:
\begin{eqnarray} 
\label{bonlength-consistency-at-every-vertex-1}
&& \rho^-_1=1/\rho^+_1, \quad (\Leftrightarrow {\rm (A1)} ), \\
\label{bonlength-consistency-at-every-vertex-2}
&& 1/\rho^+_1=\rho^+_1, \quad  \rho^-_1=1/ \rho^-_1      \quad (\Leftrightarrow {\rm (A2)} ),
\end{eqnarray} 
then we have $\rho^+_1=\rho^-_1=1$ (vertex $1$ for simplicity). In this case, we have a trivial model because  $g_{ab}\!=\!\delta_{ab}$.

The discrete expression of the induced metric $g_{ab}\!=\!\partial_a {\bf r}\cdot\partial_b {\bf r} $ is given by $g_{ab}\!=\!\left(  
       \begin{array}{@{\,}cc}
        \ell_{12}^2 &  \vec \ell_{12}\cdot \vec \ell_{13} \\
        \vec \ell_{12}\cdot \vec \ell_{13} &  \ell_{13}^2 
       \end{array} 
       \\  \right)$, 
which is defined on triangle $123$ in ${\bf R}^3$ with the local coordinate origin is at ${\bf r}_1$ (see Fig.\ref{fig-2}(b)). This $g_{ab}$  is not of the form $\left(  
       \begin{array}{@{\,}cc}
        E &  0 \\
        0 &  G 
       \end{array} 
       \\  \right)$, and for this reason the induced metric model is out of the scope of {\it Definition 1}.   However,  it is easy to see that the induced metric model satisfies (A1)--(A3). Indeed, the bond length of this model is just the Euclidean length of bond $12$ in ${\bf R}^3$. Other conditions are also easy to confirm.  
       
%%%%%%%%%%%%%%%%%%%%%%%%%%%%%%%%%%%%%%%%%%
\subsection{Orientation symmetric model}
\begin{figure}[H]
\centering
\includegraphics[width=12cm]{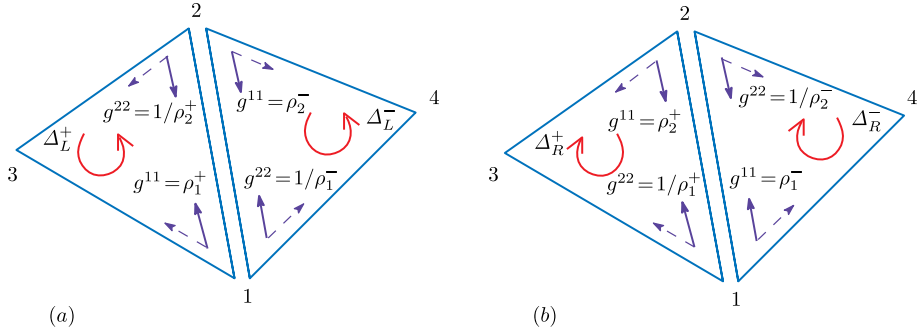}
\caption{(a) Two neighboring triangles ${\it \Delta}^\pm_L$  and elements of the inverse metric $g^{ab}$ for $S_1$ and $S_2$, and (b) the inverted triangles  ${\it \Delta}^\pm_R$ (inside view).  }
\label{fig-4}
\end{figure}

%%%%%%%%%%%%%%%%%%%%%%%%%%%%%%%%%%%%%%%%%%
The discrete model is defined by Hamiltonian  in Eq. (\ref{Disc-Eneg-model-3}), where  $g_{ab}$ is a coordinate dependent metric. Therefore, the Hamiltonian depends on the local coordinates on $M$, and it also depends on the orientation of $M$. For this reason, we define the notion of {\it orientation symmetric/asymmetric model} defined on surfaces with  ${\it \Delta}_L$.  This simply means that Hamiltonian of Eq. (\ref{Disc-Eneg-model-3}) can be used for a model in which the partition function allows the surface inversion process. Indeed, a property of the model corresponding to symmetries in Hamiltonian can be discussed without referencing the partition function in general. Thus, Hamiltonian  is called {\it orientation symmetric} if it is invariant under the surface inversion  in Eq. (\ref{inversion}), for example, for any configuration of ${\bf r}$,  and we also have:
\begin {Definition}
A discrete surface model is called orientation symmetric if the Hamiltonian is orientation symmetric.
\end {Definition}
In the Hamiltonian of Eq. (\ref{Disc-Eneg-model-3}), the quantities  $\gamma_{ij}$ and $\kappa_{ij}$ in $S_1$ and $S_2$ depend on the surface orientation. Thus, the condition for that the Hamiltonian is orientation symmetric  is as follows:
\begin{eqnarray} 
\label{orient-symmetry-Hamiltonian}
&& 1/\rho^-_1+\rho^-_2+1/\rho^+_2+\rho^+_1= 1/\rho^-_2+\rho^-_1+1/\rho^+_1+\rho^+_2
\end{eqnarray} 
for all bonds $12$ and ${\it \Delta}^\pm$. Indeed, the Gaussian bond potential $S_1(\ell_{12})$ of bond $12$ is given by  $S_1(\ell_{12})\!=\!(1/4)\left(1/\rho^-_1\!+\!\rho^-_2\!+\!1/\rho^+_2\!+\!\rho^+_1\right)\ell_{12}^2$ (Fig.\ref{fig-4}(a)), while on the inverted triangles the corresponding quantity $\bar S_1(\ell_{12})$ is given by  $\bar S_1(\ell_{12})\!=\!(1/4)\left(1/\rho^-_2\!+\!\rho^-_1\!+\!1/\rho^+_1\!+\!\rho^+_2\right)\ell_{12}^2$. These  $S_1(\ell_{12})$ and $\bar S_1(\ell_{12})$ are obtained by using the following expression for the inverse metric: 
\begin{equation}
\label{inverse_metric}
g^{ab}=g_{ab}^{-1}=\left(  
       \begin{array}{@{\,}cc}
        \rho &  0 \\
        0 &  1/\rho
       \end{array} 
       \\ 
 \right).
\end{equation} 
Thus, from the equation $S_1(\ell_{12})\!=\!\bar S_1(\ell_{12})$ for any bond $12$, which is the condition for  $S_1$ to be  orientation symmetric, we have Eq. (\ref{orient-symmetry-Hamiltonian}). We should note that from the condition  $S_2({\bf n}^+\cdot{\bf n}^-)\!=\!\bar S_2({\bf n}^+\cdot{\bf n}^-)$ for the bending energy $S_2$ the same equation as Eq. (\ref{orient-symmetry-Hamiltonian}) is obtained.
%=====================================
%=====================================

\begin{Remark}We have the following remarks:
\begin{enumerate}[leftmargin=1cm,labelsep=3mm]
\item[{\rm (a)}]	All non-trivial models are orientation asymmetric
\item[{\rm (b)}]	All orientation asymmetric models are ill-defined
\end{enumerate}
\end{Remark}
\begin{proof}[Proof of Remark 1]  (a) The inverse metric $g^{ab}$ of a non-trivial model is given by Eq. (\ref{inverse_metric}),  and therefore, it is easy to see that there exists a bond $12$ such that $S_1(\ell_{12})\!\not=\!\bar S_1(\ell_{12})$.  Indeed, we can choose $\rho$'s such that Eq. (\ref{orient-symmetry-Hamiltonian}) is not satisfied. This inequality $S_1(\ell_{12})\!\not=\!\bar S_1(\ell_{12})$ implies that the condition in Eq. (\ref{orient-symmetry-Hamiltonian}) is not satisfied and that the model is orientation asymmetric.  (b)  $\Leftrightarrow$  {\it  All well-defined models are orientation symmetric}, which can be proved as follows: If the model is well-defined, then Eqs. (\ref{expression-definition-1-1}) and (\ref{expression-definition-1-2}) are satisfied. Then, it is easy to see that Eq. (\ref{orient-symmetry-Hamiltonian}) is satisfied. This implies that the model is orientation symmetric. 
\end{proof}

From Remark 1, it is straightforward to prove the following theorem:
\begin{Theorem}
All non-trivial models are ill-defined.
\end{Theorem}

Here, we should clarify how well-defined models are different from the model with Euclidean metric $\delta_{ab}$.  This problem is rephrased such that what type of $\rho$ is allowed for a well-defined model. The answer is as follows: 
\begin{Remark}We have the following remarks:
\begin{enumerate}[leftmargin=1cm,labelsep=3mm]
\item[{\rm (a)}] The function $\rho_i$ of  any well-defined model satisfies
\begin{eqnarray}
\label{trivial-relation-1}
1/\rho_i+\rho_i=a(={\rm const}),
\end{eqnarray}
where the constant $a$ depends on neither vertex $i$ nor triangle ${\it \Delta}$.
\item[{\rm (b)}] There are two possible $\rho$s, which are solutions of Eq. {\rm (\ref{trivial-relation-1}):}
\begin{eqnarray}
\label{trivial-rho}
\rho_{\pm}=\frac{1}{2}\left(a\pm \sqrt{a^2-4}\right)
\end{eqnarray}
\item[{\rm (c)}] If Eq. {\rm (\ref{trivial-relation-1})} is satisfied, then  the model is trivial.
\end{enumerate}
\end{Remark}

\begin{proof}[Proof of Remark 2]  (a) A well-defined model satisfies Eqs. (\ref{expression-definition-1-1}) and (\ref{expression-definition-1-2}). Multiplying  both sides of the first equation in Eq. (\ref{expression-definition-1-2}) by $\rho_1^-\rho_2^-(>\!0)$, we have $\rho_1^-\left(\rho_1^-\rho_2^-\!+\!1\right)\!=\!\rho_2^-\left(\rho_1^-\rho_2^-\!+\!1\right)$, and therefore $\rho_1^-\!=\!\rho_2^-(=\rho^-)$.  It is also easy to see that $\rho_1^+\!=\!\rho_2^+(=\rho^+)$ from the second equation in Eq. (\ref{expression-definition-1-2}). Therefore, using these two equations and Eq. (\ref{expression-definition-1-1}), we have $\rho_1^-\!+\!1/\rho_1^-\!=\!\rho_1^+\!+\!1/\rho_1^+$. This implies that the combination $\rho^-\!+\!1/\rho^-$ is independent of the vertex and triangle, and thus Eq. (\ref{trivial-relation-1}) is proved.  (b) It is easy to see that $\rho_{\pm}\!=\!\left(a\!\pm\! \sqrt{a^2\!-\!4}\right)/2$, ($a\!\geq\! 2$) from Eq. (\ref{trivial-relation-1}).  (c)  Indeed, using Eq. (\ref{trivial-relation-1}), we have $\gamma_{ij}\!=\!\kappa_{ji}\!=\!(1/4)\left(\rho^+\!+\!1/\rho^+\!+\!\rho^-\!+\!1/\rho^-\right)\!=\!a/2$, and therefore  $S_1(a)\!=\!\sum_{ij}\gamma_{ij}\ell_{ij}^2\!=\!(a/2)\sum_{ij}\ell_{ij}^2$ and  $S_2(a)\!=\!\sum_{ij}\kappa_{ij}\left(1\!-\!{\bf n}_i\cdot{\bf n}_j\right)\!=\!(a/2)\sum_{ij}\left(1\!-\!{\bf n}_i\cdot{\bf n}_j\right)$.  
 \end{proof}

 It follows from Remark 2(a) that the model in Ref. \cite{Koibuchi-Pol2016} is ill-defined (in the context of HP model). In fact, the metric function assumed in the model of Ref. \cite{Koibuchi-Pol2016} does not satisfy  Eq. (\ref{trivial-relation-1}).  The metric corresponding to Remark 2(b) shows examples of metric for the trivial model, which is defined by Definition 1. More explicitly,
$\left(  
      \begin{array}{@{\,}cc}
        1/\rho_+ & 0 \\
        0 &  \rho_+
       \end{array} 
       \\ 
       \right)$ and 
 $\left(  
     \begin{array}{@{\,}cc}
        1/\rho_- & 0 \\
        0 &  \rho_-
     \end{array} 
      \\ 
\right)$ make the model trivial. 
 The metrices 
 $\left(  
      \begin{array}{@{\,}cc}
        1/\rho_+ & 0 \\
        0 &  \rho_-
       \end{array} 
       \\ 
       \right)$ and 
$\left(  
     \begin{array}{@{\,}cc}
        1/\rho_- & 0 \\
        0 &  \rho_+
     \end{array} 
      \\ 
\right)$ are conformally equivalent to $\delta_{ab}$, because $\rho_+ \rho_-\!=\!1$, and therefore these also make the model trivial.  We should remark that Remarks 2(a) and 2(c) also prove Theorem 1. 

Note also that if a model is well-defined and orientation symmetric in the sense of Definitions 2 and 3 then inverted triangles ${\it \Delta}_R$ need not to be included in the lattice configuration. However, from Theorem 1 the model introduced in Eq. (\ref{Disc-Eneg-model-3}) is orientation asymmetric, and this model turns to be well-defined if it is treated as an FG model.  Therefore the inverted triangles ${\it \Delta}_R$ should be included as a representation configuration of the model of Eq. (\ref{Disc-Eneg-model-3}) if it is understood as a well-defined model. For this reason, we have to extend the FG model introduced in Ref. \cite{Koibuchi-Sekino-2014PhysicaA} such that the Hamiltonian has values on both ${\it \Delta}_L$ and ${\it \Delta}_R$.
  
%%%%%%%%%%%%%%%%%%%%%%%%%%%%%%%%%%%%%%%%%%
\section{Finsler geometry modeling}\label{FG-model}
%%%%%%%%%%%%%%%%%%%%%%%%%%%%%%%%%%%%%%%%%%
%%%%%%%%%%%%%%%%%%%%%%%%%%%%%%%%%%%%%%%%%%
\subsection{Finsler geometry model}
%%%%%%%%%%%%%%%%%%%%%%%%%%%%%%%%%%%%%%%%%%
As we have demonstrated in the previous subsection, all non-trivial surface models ($\Leftrightarrow$ either $\gamma_{ij}$ or $\kappa_{ij}$ depends on $ij$) are ill-defined.   The reason why this unsatisfactory result is obtained is because the bond length should not be direction dependent for any well-defined models (see Definition 2). To make this ill-defined models meaningful, we introduce the notion of Finsler geometry, where length unit is allowed to be dependent on the direction. In the context of Finsler geometry modeling, Theorem 1 does not hold. The problem is whether or not the above mentioned ill-defined model (in Section \ref{discrete-model}) is fitted in Finsler geometry modeling.

Let  ${\it \Delta}_{L,R}$ be triangles in $M$, and $x\!=\!(x_1,x_2)$ be a local coordinate on   ${\it \Delta}_{L,R}$, where the coordinate origin is at vertex $1$.  Let  $y\!=\!(y_1,y_2)$ be defined by $y_i\!=\!dx_i/dt,\; (i\!=\!1,2)$, where $t$ is a parameter that increases toward the positive direction of the axes. It is also assumed that a positive parameter $v_{ij}$ is defined on the axis from vertex $i$ to vertex $j$, where $v_{ij}\!\not=\!v_{ji}$ in general.

%=====================================
\begin{figure}[H]
\centering
\includegraphics[width=13cm]{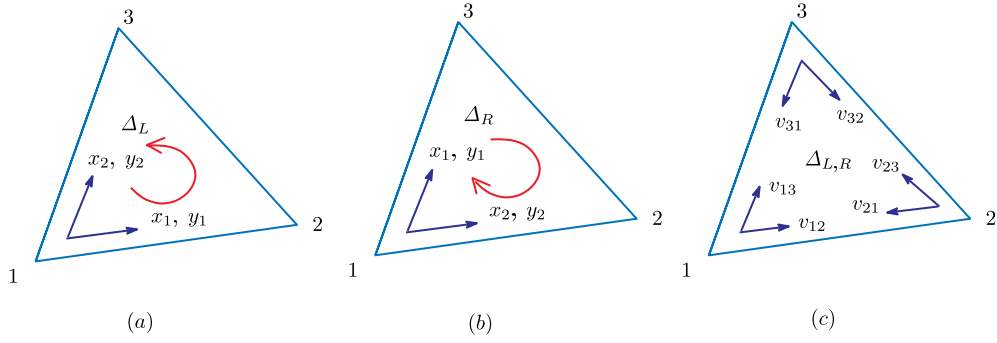}
\caption{ A triangle $123$ with the local coordinate axis $x_i$ and the tangent vector component $y_i(=\!\dot x_i)$ at vertex $1$ on (a) left-handed triangle  ${\it \Delta}_{L}$ and  (b) right-handed triangle  ${\it \Delta}_{R}$, and (c) three possible local coordinates on triangle ${\it \Delta}_{L,R}$, and positive number $v_{ij}$ assigned along the bond $ij$.  }
\label{fig-5}
\end{figure}
%=====================================
Discrete Finsler functions on triangles ${\it \Delta}_{L,R}$ in $M$ are defined by (Figs. \ref{fig-5}(a),(b))
\begin{eqnarray}
\label{Discrete-FG-func-1}
L_{{\it \Delta}_L}(x,y)= \left\{ \begin{array}{@{\,}ll}
                 y_1/v_{12}  &  ({\rm on}\; x_1 \; {\rm axis})  \\
                 y_2/v_{13} &  ({\rm on}\; x_2 \; {\rm axis}) 
                  \end{array} 
                   \right., \quad
  L_{{\it \Delta}_R}(x,y)= \left\{ \begin{array}{@{\,}ll}
                 y_1/v_{13}  &  ({\rm on}\; x_1 \; {\rm axis})  \\
                 y_2/v_{12} &  ({\rm on}\; x_2 \; {\rm axis}) 
                  \end{array} 
                   \right., 
\end{eqnarray} 
which can also be written as the bilinear forms
\begin{eqnarray}
\label{Discrete-FG-func-12}
L_{{\it \Delta}_L}^2(x,y)= {v_{12}^{-2}y_1^2 + v_{13}^{-2}y_2^2},\quad
L_{{\it \Delta}_R}^2(x,y)= {v_{13}^{-2}y_1^2 + v_{12}^{-2}y_2^2}.
\end{eqnarray} 
From these expressions, we have the metric functions $g_{ab,L}$ on ${\it \Delta}_{L}$ and and $g_{ab,R}(x)$ on ${\it \Delta}_{R}$, such that 
\begin{equation}
\label{metric-from-L}
g_{ab,L}(x)=\frac{1}{2}\frac{\partial L_{{\it \Delta}_L}^2(x,y)}{\partial y_a\partial y_b}=\left(  
       \begin{array}{@{\,}cc}
        v_{12}^{-2} &  0 \\
        0 &  v_{13}^{-2} 
       \end{array} 
       \\ 
 \right), \quad
 g_{ab,R}(x)=\frac{1}{2}\frac{\partial L_{{\it \Delta}_R}^2(x,y)}{\partial y_a\partial y_b}=\left(  
       \begin{array}{@{\,}cc}
        v_{13}^{-2} &  0 \\
        0 &  v_{12}^{-2} 
       \end{array} 
       \\ 
 \right).
\end{equation} 
In general, $g_{ab}$ is a function with respect to $x$ and $y$, however, $g_{ab, LR}$  in Eq. (\ref{metric-from-L}) only depends on the local coordinate $x$ and it is independent of $y$. 

Using the metric $g_{ab, LR}$ in Eq. (\ref{metric-from-L}) and summing over all possible coordinate origins on  triangle ${\it \Delta}_{L,R}$,  just the same as in Eq. (\ref{Disc-Eneg-model-2}), we have the discrete Hamiltonian such that (see Fig. \ref{fig-2}(b))
\begin{eqnarray}
\label{Discrete-FG-func-L-1}
\begin{split}
&S_1=\sum_{\it \Delta}\left(\gamma_{12}\ell_{12}^2+\gamma_{23}\ell_{23}^2+\gamma_{31}\ell_{31}^2\right),\\
&S_2=\sum_{\it \Delta}\left[\kappa_{12}\left(1-{\bf n}_0\cdot{\bf n}_3\right)+\kappa_{23}\left(1-{\bf n}_0\cdot{\bf n}_1\right)+\kappa_{31}\left(1-{\bf n}_0\cdot{\bf n}_2\right)\right],\\
&\gamma_{12}=\frac{v_{12}}{v_{13}}+\frac{v_{21}}{v_{23}}, \quad\gamma_{23}=\frac{v_{23}}{v_{21}}+\frac{v_{32}}{v_{31}}, \quad\gamma_{31}=\frac{v_{31}}{v_{32}}+\frac{v_{13}}{v_{12}}, \\
&\kappa_{12}=\frac{v_{13}}{v_{12}}+\frac{v_{23}}{v_{21}}, \quad\kappa_{23}=\frac{v_{21}}{v_{23}}+\frac{v_{31}}{v_{32}}, \quad\kappa_{31}=\frac{v_{32}}{v_{31}}+\frac{v_{12}}{v_{13}}.
\end{split}
\end{eqnarray}

\begin{figure}[H]
\centering
\includegraphics[width=12cm]{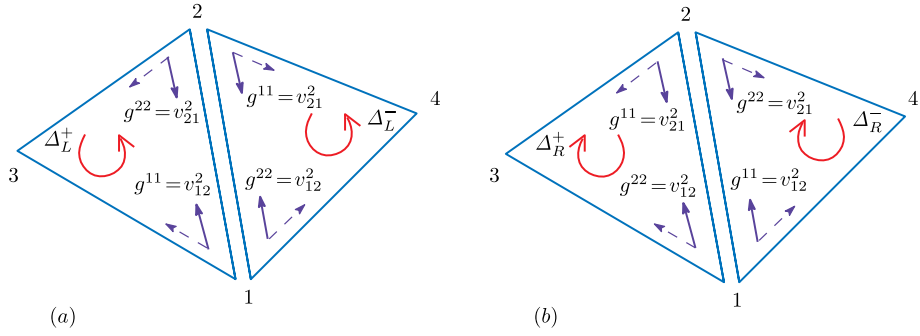}
\caption{Two of four possible combinations of triangles ${\it \Delta}^+_{L,R}$ and ${\it \Delta}^-_{L,R}$, which share  bond $12$:  (a) ${\it \Delta}^+_L$ and ${\it \Delta}^-_L$ and (b) the inverted triangles ${\it \Delta}^+_R$ and ${\it \Delta}^-_R$  (inside view) of those in (a).  Elements of the inverse metric $g^{ab}$ are given by $v_{12}^2$ and  $v_{21}^2$, which are defined on bond $12$. }
\label{fig-6}
\end{figure}

The sum over triangles $\sum_{\it \Delta}$ in $S_1$ and  $S_2$ can also be expressed by the sum over bonds with a numerical factor $1/4$. Thus, we have 
\begin{eqnarray}
\label{Discrete-FG-func-L-2}
\begin{split}
&S=S_1+\kappa S_2, \\
&S_1=\frac{1}{4}\sum_{ij}\left(\gamma_{ij}^++\gamma_{ij}^-\right)\ell_{ij}^2,\quad S_2=\frac{1}{4}\sum_{ij}\left(\kappa_{ij}^++\kappa_{ij}^-\right)\left(1-{\bf n}^+\cdot{\bf n}^-\right),\\
&\gamma_{12}^+=\frac{v_{12}}{v_{13}}+\frac{v_{21}}{v_{23}}, \quad \gamma_{12}^-=\frac{v_{12}}{v_{14}}+\frac{v_{21}}{v_{24}}, \quad
\kappa_{12}^+=\frac{v_{13}}{v_{12}}+\frac{v_{23}}{v_{21}}, \quad \kappa_{12}^-=\frac{v_{14}}{v_{12}}+\frac{v_{24}}{v_{21}},
\end{split}
\end{eqnarray} 
 where  $\gamma_{12}^\pm$ and $\kappa_{12}^\pm$  are concrete examples of $\gamma_{ij}^\pm$ and $\kappa_{ij}^\pm$ for bond $12$ (see Fig. \ref{fig-5}(c)). The symbol $\pm$ denotes that  $\gamma_{ij}$ and $\kappa_{ij}$ defined  on the triangles ${\it \Delta}^\pm_{L,R}$ which share the bond $ij$ (Figs. \ref{fig-6} (a),(b)). 
%=====================================
%=====================================
 
  If the coefficients $\gamma_{ij}^\pm$ and $\kappa_{ij}^\pm$ are defined by the quantities, which are defined on vertices $i$ and $j$ or on bond $ij$, just like those in Eq. (\ref{Discrete-FG-func-L-2}), then these coefficients become independent of  the orientation of the triangles. Therefore, we have $\gamma_{ij,L}^\pm\!=\!\gamma_{ij,R}^\pm\left(=\!\gamma_{ij}^\pm\right)$, and $\kappa_{ij,L}^\pm\!=\!\kappa_{ij,R}^\pm\left(=\!\kappa_{ij}^\pm\right)$,  and  therefore the model is orientation symmetric. In this case,  we have $\gamma_{ij}^+\!+\!\gamma_{ij}^-\!=\!\gamma_{ji}^+\!+\!\gamma_{ji}^-\left(=\!4\gamma_{ij}\right)$, and $\kappa_{ij}^+\!+\!\kappa_{ij}^-\!=\!\kappa_{ji}^+\!+\!\kappa_{ji}^-\left(=\!4\kappa_{ij}\right)$. On the contrary, if $\gamma_{ij}^\pm$ and $\kappa_{ij}^\pm$ depend on ${\it \Delta}_{L,R}$, then  the model is orientation asymmetric. In this case, we have $\gamma_{ij,LR}^+\!+\!\gamma_{ij,LR}^-\!\not=\!\gamma_{ji,LR}^+\!+\!\gamma_{ji,LR}^- (\Leftrightarrow \!\gamma_{ij,L}^+\!+\!\gamma_{ij,L}^-\!\not=\!\gamma_{ji,L}^+\!+\!\gamma_{ji,L}^-, \cdots)$ and $\kappa_{ij,LR}^+\!+\!\kappa_{ij,LR}^-\!\not=\!\kappa_{ji,LR}^+\!+\!\kappa_{ji,LR}^-$ in general. It is also easy to see that $\gamma_{ij,LR}^+\!+\!\gamma_{ij,RL}^-\!\not=\!\gamma_{ji,LR}^+\!+\!\gamma_{ji,RL}^- (\Leftrightarrow \!\gamma_{ij,L}^+\!+\!\gamma_{ij,R}^-\!\not=\!\gamma_{ji,L}^+\!+\!\gamma_{ji,R}^-, \cdots)$ and $\kappa_{ij,LR}^+\!+\!\kappa_{ij,RL}^-\!\not=\!\kappa_{ji,LR}^+\!+\!\kappa_{ji,RL}^-$.
Such an orientation asymmetric FG model will be studied in the following subsection.

Finally in this subsection, we emphasize a difference between the models defined by Eqs. (\ref{Discrete-FG-func-L-2}) and (\ref{Disc-Eneg-model-3}). In fact, the expressions of $S_1$ and  $S_2$ in Eq. (\ref{Discrete-FG-func-L-2})  are different from those in Eq. (\ref{Disc-Eneg-model-3}). This difference comes from the fact that $S_1$ and  $S_2$ in Eq. (\ref{Disc-Eneg-model-3}) are simply obtained by discretization of an ordinary HP surface model with a non-Euclidean metric. More explicitly, we have the following facts: 
 %\begin{enumerate}
  (i) Not only ${\it \Delta}^\pm_{L}$ but also  ${\it \Delta}^\pm_{R}$ is assumed to define $S_1$ and  $S_2$ in Eq. (\ref{Discrete-FG-func-L-2}) , while only ${\it \Delta}^\pm_{L}$ is assumed to define those in  Eq. (\ref{Disc-Eneg-model-3}). 
 (ii) Finsler function is assumed to define $S_1$ and  $S_2$  in Eq. (\ref{Discrete-FG-func-L-2}),  while it is not assumed to define those in  Eq. (\ref{Disc-Eneg-model-3}).  
 %\end{enumerate}
Therefore, mainly from the latter fact (ii), it is still unclear whether the model defined by Eq.  (\ref{Disc-Eneg-model-3}) can be called an FG model or not. For this reason, the model defined by Eq.  (\ref{Disc-Eneg-model-3}) still remains  ill-defined, although the Hamiltonian in Eq.  (\ref{Disc-Eneg-model-3})  is very close to the one in Eq. (\ref{Discrete-FG-func-L-2}).

%%%%%%%%%%%%%%%%%%%%%%%%%%%%%%%%%%%%%%%%%%
\subsection{Orientation asymmetric Finsler geometry model}
%%%%%%%%%%%%%%%%%%%%%%%%%%%%%%%%%%%%%%%%%%
As we have discussed in the previous subsection, FG model in Ref. \cite{Koibuchi-Sekino-2014PhysicaA} is extended such that inverted triangles are included in the lattices.  The triangulated lattices are composed of both  ${\it \Delta}_L$ and ${\it \Delta}_R$, where  ${\it \Delta}_R$ corresponds to an inverted part of surface (Fig. \ref{fig-1}(d)).  On these triangles ${\it \Delta}_L$ and ${\it \Delta}_R$, the coefficients $\gamma_{ij}^\pm$ and $\kappa_{ij}^\pm$ of $S_1$ and $S_2$ are defined. Therefore, the orientation asymmetric states are in general allowed in the configurations of the FG model. In this subsection, we show that the ill-defined model constructed in the previous Section by Eq.  (\ref{Disc-Eneg-model-3}) turns to be a well-defined model in the context of FG modeling.

By comparing $g_{ab}$ in Eq. (\ref{metric-from-L}) and $g_{ab}$ in Eq. (\ref{org_cont_metric}), we have the following correspondence between the parameters $v_{12}, v_{13},\cdots, v_{42}$  and the functions $\rho_1^\pm, \rho_2^\pm, \rho_3^\pm$ on  ${\it \Delta}^\pm_L$ (see  Figs. \ref{fig-4}(a), \ref{fig-5}(c) and \ref{fig-6}(a)):
\begin{eqnarray}
\label{v-vs-rho-L}
\begin{split}
&v_{12}^{-2}=1/\rho_1^+, \quad v_{13}^{-2}= \rho_1^+, \quad v_{23}^{-2}=1/\rho_2^+, \quad v_{21}^{-2}= \rho_2^+, \quad
v_{31}^{-2}=1/\rho_3^+, \quad v_{32}^{-2}= \rho_3^+,\quad ({\rm on}\; {\it \Delta}^+_L),\\
&v_{12}^{-2}=1/\rho_1^-, \quad v_{14}^{-2}= \rho_1^-, \quad v_{24}^{-2}=1/\rho_2^-, \quad v_{21}^{-2}= \rho_2^-, \quad
v_{41}^{-2}=1/\rho_3^-, \quad v_{42}^{-2}= \rho_3^-,\quad ({\rm on}\; {\it \Delta}^-_L).
\end{split}
\end{eqnarray} 
The symbol $\rho_i^+$ is a function on triangle ${\it \Delta}^+_L$ for the metric in Eq. (\ref{org_cont_metric}) when the local coordinate is at vertex $i(=1,2,3)$. We also have a contribution from ${\it \Delta}^\pm_R$:
\begin{eqnarray}
\label{v-vs-rho-R}
\begin{split}
&v_{12}^{-2}=\rho_1^+, \quad v_{13}^{-2}= 1/\rho_1^+, \quad v_{23}^{-2}=\rho_2^+, \quad v_{21}^{-2}= 1/\rho_2^+, \quad
v_{31}^{-2}=\rho_3^+, \quad v_{32}^{-2}= 1/\rho_3^+,\quad ({\rm on}\; {\it \Delta}^+_R),\\
&v_{12}^{-2}=\rho_1^-, \quad v_{14}^{-2}= 1/\rho_1^-, \quad v_{24}^{-2}=\rho_2^-, \quad v_{21}^{-2}= 1/\rho_2^-, \quad
v_{41}^{-2}=\rho_3^-, \quad v_{42}^{-2}= 1/\rho_3^-,\quad ({\rm on}\; {\it \Delta}^-_R).
\end{split}
\end{eqnarray} 
By inserting these expressions into $\gamma_{ij}$ and $\kappa_{ij}$ in Eq. (\ref{Discrete-FG-func-L-2}) ($v_{41}^{-2}$ and $v_{42}^{-2}$ in Eqs. (\ref{v-vs-rho-L}) and (\ref{v-vs-rho-R}) are not included in the list  below), we have 
\begin{eqnarray}
\label{gamma-kappa-FG-func-L}
\begin{split}
&\gamma_{12}^+={\rho_1^+}+{1}/{\rho_2^+}, \quad\gamma_{23}^+={\rho_2^+}+{1}/{\rho_3^+}, \quad\gamma_{31}^+={\rho_3^+}+{1}/{\rho_1^+}, \quad ({\rm on}\; {\it \Delta}^+_L), \\
&\gamma_{12}^-={\rho_2^-}+{1}/{\rho_1^-}, \quad\gamma_{23}^-={\rho_3^-}+{1}/{\rho_2^-}, \quad\gamma_{31}^-={\rho_1^-}+{1}/{\rho_3^-}, \quad ({\rm on}\; {\it \Delta}^-_L), \\
&\kappa_{12}^+={\rho_2^+}+1/{\rho_1^+}, \quad\kappa_{23}^+={\rho_3^+}+{1}/{\rho_2^+}, \quad\kappa_{31}^+={\rho_1^+}+{1}/{\rho_3^+},\quad ({\rm on}\; {\it \Delta}^+_L),\\
&\kappa_{12}^-={\rho_1^-}+{1}/{\rho_2^-}, \quad\kappa_{23}^+={\rho_2^-}+{1}/{\rho_3^-}, \quad\kappa_{31}^-={\rho_3^-}+{1}/{\rho_1^-},\quad ({\rm on}\; {\it \Delta}^-_L). 
\end{split}
\end{eqnarray} 
The expressions of $\gamma_{ij}^\pm$ and $\kappa_{ij}^\pm$  on ${\it \Delta}^\pm_R$ are obtained by replacing $\rho$ with $1/\rho$ in the expressions in Eq. (\ref{gamma-kappa-FG-func-L}). We find from Eq. (\ref{gamma-kappa-FG-func-L}) that the coefficients $\gamma_{ij}^\pm$ and $\kappa_{ij}^\pm$ can also be written more simply by using the suffices $ij$, which will be presented below. 

To incorporate two types of triangles ${\it \Delta}_{L,R}$ into the lattice configurations, which are dynamically updated in the partition function, we need a new variable corresponding to these  ${\it \Delta}_{L,R}$. Thus, we introduce a new dynamical variable $\chi$, which is defined on triangles ${\it \Delta}$ and has values in ${\bf Z}_2$ just like $\sigma$ in Eq. (\ref{Part-Func-flu}) to represent the surface orientation:
\begin{eqnarray}
\label{orientation-variable}
\chi\left({\it \Delta}\right)= \left\{ \begin{array}{@{\,}ll}
                 1  &  ({\it \Delta}={\it \Delta}_L)  \\
                 -1 &  ({\it \Delta}={\it \Delta}_R) 
                  \end{array} 
                   \right.. 
\end{eqnarray} 
If $\chi_i\left(=\!\chi({\it \Delta}_i)\right)\!=\!-1$ is satisfied for all triangles ${\it \Delta}_i$, then the surface is understood as it is completely inverted.  In contrast, mixed states, where the value of $\chi_i$ is not uniform, are understood as  a partly inverted membrane (see Fig. \ref{fig-1}(d)). This implies that actual intersections like the one in Fig. \ref{fig-1}(b) are not necessarily implemented in the model. If such intersections must be taken into consideration in the numerical simulation, it will be very time consuming, because every step for the vertex move should be checked to monitor how the lattice intersects. More than that the simulation is time consuming, as mentioned in the previous section real physical membranes are expected to undergo inversion by pore formation without self-intersection.  

By this new variable  $\chi_i$ in Eq. (\ref{orientation-variable}),  the FG model introduced in \cite{Koibuchi-Sekino-2014PhysicaA} is  extended such that the inverted surface states are included in the surface configurations. Indeed, for any given configuration, its inverted configuration by Eq. (\ref{inversion}) is included in the configurations, because the inverted configuration is obtained by the transformation $\chi_i \to -\chi_i$ for all $i$ and with suitable translation and deformation of ${\bf r}$.   In this new model, the triangulated surfaces are composed of both  ${\it \Delta}_L$ and ${\it \Delta}_R$, where  the triangles ${\it \Delta}_R$ correspond to an inverted part of surface like the one in Fig. \ref{fig-1}(d). The coefficients $\gamma_{ij}$ and $\kappa_{ij}$ of $S_1$ and $S_2$ are defined on not only ${\it \Delta}_L$ but also ${\it \Delta}_R$. Therefore, the orientation asymmetric states are naturally expected in the configurations of the new model.

The variable $\chi$ has values in ${\bf Z}_2$ just like $\sigma$ in the energy $S_0$ of Eq.(\ref{Part-Func-flu}), however, the role of $\chi$ is different from that of $\sigma$. The variable $\sigma$ plays a role for defining the functions $\rho_i$ of the metric $g_{ab}$. In the context of the modeling in this paper, $\rho$ is determined  independently of the surface orientation $\chi$.  As mentioned in the end of Section \ref{discretization}, $S_0$ is not included in the Hamiltonian introduced below although the role of $S_0$ is completely different from that of $S_3$.

By including the partition function, we finally have
\begin{eqnarray}
\label{Asymm-FG-model}
\begin{split}
&Z(\zeta,\kappa) =  \sum_\chi \sum_{\mathcal T}\int^\prime \prod _{i=1}^{N} d {\bf r}_i \exp\left[-S({\bf r},\chi)\right],\quad
S=S_1+\kappa S_2+\zeta S_3, \\
&S_1=\frac{1}{4}\sum_{ij}\left(\gamma_{ij}^++\gamma_{ij}^-\right)\ell_{ij}^2,\quad S_2=\frac{1}{4}\sum_{ij}\left(\kappa_{ij}^++\kappa_{ij}^-\right)\left(1-{\bf n}^+\cdot{\bf n}^-\right),\\
&S_3=\sum_{\pm}\left(1-\chi^+\cdot\chi^-\right),\quad \left(\chi^\pm\in \{1,-1\}\right),\\
&\gamma_{ij}^+=\left\{ \begin{array}{@{\,}ll}
                 \rho^+_{i}+{1}/{\rho^+_{j}} &  \left(\chi({\it \Delta}^+)=1\right) \; \\
                 1/\rho^+_{i}+{\rho^+_{j}} &  \left(\chi({\it \Delta}^+)=-1\right)\; 
                   \end{array} 
                   \right., 
                \quad \gamma_{ij}^-=\left\{ \begin{array}{@{\,}ll}
                 \rho^-_{j}+{1}/{\rho^-_{i}} &  \left(\chi({\it \Delta}^-)=1\right) \; \\
                 1/\rho^-_{j}+{\rho^-_{i}} &  \left(\chi({\it \Delta}^-)=-1\right) \; 
                   \end{array} 
                   \right., \\
&\kappa_{ij}^+=\left\{ \begin{array}{@{\,}ll}
                 \rho^+_{j}+{1}/{\rho^+_{i}} &  \left(\chi({\it \Delta}^+)=1\right) \; \\
                 1/\rho^+_{j}+{\rho^+_{i}} &  \left(\chi({\it \Delta}^+)=-1\right) \; 
                   \end{array} 
                   \right., 
                \quad \kappa_{ij}^-=\left\{ \begin{array}{@{\,}ll}
                 \rho^-_{i}+{1}/{\rho^-_{j}} &  \left(\chi({\it \Delta}^-)=1\right) \; \\
                 1/\rho^-_{i}+{\rho^-_{j}} &  \left(\chi({\it \Delta}^-)=-1\right) \; 
                   \end{array} 
                   \right.,
\end{split}
\end{eqnarray} 
where Ising model Hamiltonian $S_3$ is assumed for the variable $\chi$ with the coefficient $\zeta$. The value of $\chi^\pm \left(\in \{1,-1\}\right)$ corresponds to ${\it \Delta}^\pm_{L,R}$ as in Eq. (\ref{orientation-variable}).  For sufficiently large $\zeta$, one of the lowest energy states of $S_3$ is realized because both $S_1$ and $S_2$ are asymmetric even though $S_3$ is symmetric under the surface inversion.  Thus, we have proved that the model introduced in Eq. (\ref{Disc-Eneg-model-3}) is identified to the FG model defined by  Eq. (\ref{Discrete-FG-func-L-2}), in which the Finsler functions in Eq. (\ref{Discrete-FG-func-1}) are assumed.  We should note that Ising model Hamiltonian is not always necessary for $S_3$. Note also that this FG model in Eq. (\ref{Asymm-FG-model}) has no constraint for the well-definedness introduced in Definition 2. In this sense, this model is well-defined even though the bond length in $M$ is direction dependent.  Moreover, since the surface configuration includes inverted triangles,  this model is orientation asymmetric from Remark 1 (a). Thus, we have
\begin{Theorem} 
\label{bond-length-v}
All non-trivial models such as the one defined by Eq. (\ref{Disc-Eneg-model-3}) or Eq. (\ref{Asymm-FG-model}) are  orientation asymmetric and well-defined in the context of Finsler geometry modeling. 
\end{Theorem}

\section{Summary}\label{discussion}
%%%%%%%%%%%%%%%%%%%%%%%%%%%%%%%%%%%%%%%%%%
%%%%%%%%%%%%%%%%%%%%%%%%%%%%%%%%%%%%%%%%%%
In this paper, we confine ourselves to discrete surface models of Helfrich and Polyakov with the metric of the type $g_{ab}\!=\!\left(  
       \begin{array}{@{\,}cc}
        E & \; 0 \\
        0 & \; G 
       \end{array} 
       \\  \right)$. The discrete model is defined on dynamically triangulated surfaces in ${\bf R}^3$ , and therefore the model is aimed at describing properties of fluid membranes such as lipid bilayers. The result in this paper indicates that the surface models with this type of non-Euclidean metric are well-defined in the context of Finsler geometry (FG) modeling,  and moreover the models are orientation asymmetric in general.  Indeed, in the FG scheme for discrete surface models, length of bond of the triangles in the parameter space $M$ can be direction dependent, and no constraint is imposed on the bond length of inverted surfaces in the FG modeling. These allow us to introduce a new dynamical variable corresponding to the triangle orientation to incorporate the surface inversion process in the model.  Thus, Hamiltonian of the models with non-trivial $g_{ab}$ has values on locally inverted surface,  and for this reason the Hamiltonian becomes dependent on the surface orientation. This property is expected to be useful to study real physical membranes, which undergo surface inversion.  FG modeling for membranes and the numerical studies should be performed more extensively.

\acknowledgments{The authors acknowledge S. Bannai and M. Imada for comments and discussions. This work is supported in part by JSPS KAKENNHI Numbers 26390138 and 17K05149.}
%\acknowledgments{All sources of funding of the study should be disclosed. Please clearly indicate grants that you have received in support of your research work. Clearly state if you received funds for covering the costs to publish in open access.}

%%%%%%%%%%%%%%%%%%%%%%%%%%%%%%%%%%%%%%%%%%
%\authorcontributions{For research articles with several authors, a short paragraph specifying their individual contributions must be provided. The following statements should be used ``X.X. and Y.Y. conceived and designed the experiments; X.X. performed the experiments; X.X. and Y.Y. analyzed the data; W.W. contributed reagents/materials/analysis tools; Y.Y. wrote the paper.'' Authorship must be limited to those who have contributed substantially to the work reported.}
\authorcontributions{E.P. performed the calculations, and H.K. wrote the paper.
}
%%%%%%%%%%%%%%%%%%%%%%%%%%%%%%%%%%%%%%%%%%
\conflictofinterests{The authors declare no conflict of interest. The founding sponsors had no role in the design of the study; in the collection, analyses, or interpretation of data; in the writing of the manuscript, and in the decision to publish the results.}

%\conflictofinterests{Declare conflicts of interest or state ``The authors declare no conflict of interest.'' Authors must identify and declare any personal circumstances or interest that may be perceived as inappropriately influencing the representation or interpretation of reported research results. Any role of the funding sponsors in the design of the study; in the collection, analyses or interpretation of data; in the writing of the manuscript, or in the decision to publish the results must be declared in this section. If there is no role, please state ``The founding sponsors had no role in the design of the study; in the collection, analyses, or interpretation of data; in the writing of the manuscript, and in the decision to publish the results''.}

%%%%%%%%%%%%%%%%%%%%%%%%%%%%%%%%%%%%%%%%%%
%% optional
\abbreviations{The following abbreviations are used in this manuscript:\\

\noindent
\begin{tabular}{@{}ll}
HP & Helfrich and Polyakov\\
FG & Finsler geometry \\
FC & Fixed connectivity \\
DT & Dynamically triangulated
%MDPI & Multidisciplinary Digital Publishing Institute\\
%DOAJ & Directory of open access journals\\
%TLA & Three letter acronym\\
%LD & linear dichroism
\end{tabular}}

%%%%%%%%%%%%%%%%%%%%%%%%%%%%%%%%%%%%%%%%%%
%% optional
%\appendixtitles{no} %Leave argument "no" if all appendix headings stay EMPTY (then no dot is printed after "Appendix A"). If the appendix sections contain a heading then change the argument to "yes".
%\appendixsections{multiple} %Leave argument "multiple" if there are multiple sections. Then a counter is printed ("Appendix A?). If there is only one appendix section then change the argument to ?one? and no counter is printed (?Appendix?).
%\appendix
%\section{}
%The appendix is an optional section that can contain details and data supplemental to the main text. For example, explanations of experimental details that would disrupt the flow of the main text, but nonetheless remain crucial to understanding and reproducing the research shown; figures of replicates for experiments of which representative data is shown in the main text can be added here if brief, or as Supplementary data. Mathemtaical proofs of results not central to the paper can be added as an appendix.

%\section{}
%All appendix sections must be cited in the main text. In the appendixes, Figures, Tables, etc. should be labeled starting with `A', e.g., Figure A1, Figure A2, etc.

%%%%%%%%%%%%%%%%%%%%%%%%%%%%%%%%%%%%%%%%%%
% Citations and References in Supplementary files are permitted provided that they also appear in the reference list here.
\bibliographystyle{mdpi}

%=====================================
% References, variant A: internal bibliography
%=====================================
\renewcommand\bibname{References}

%=====================================
% References, variant B: external bibliography
%=====================================
%\bibliography{your_external_BibTeX_file}

%%%%%%%%%%%%%%%%%%%%%%%%%%%%%%%%%%%%%%%%%%
%% optional
%\sampleavailability{Samples of the compounds ...... are available from the authors.}

%%%%%%%%%%%%%%%%%%%%%%%%%%%%%%%%%%%%%%%%%%
\end{document}